\documentclass{PoS}
\usepackage{url}
\usepackage{float}

\title{Quantum 3: Learning QCD through Intuitive Play}

\ShortTitle{Quantum 3: Learning QCD through Intuitive Play
}

\author{\speaker{Tristan \"Ozkan} \\
        Department of Computer Science, Michigan State University, East Lansing, MI 48824\\
        E-mail: \email{ozkantri@msu.edu}}
\author{Huey-Wen Lin \\
Department of Physics and Astronomy \& Computational Mathematics,
  Science and Engineering, Michigan State University, East Lansing, MI 48824 \\
  E-mail: \email{hwlin@pa.msu.edu}}

\abstract{

There is a nationwide drive to get more girls into physics and coding, and some educators believe gaming could be a way to get girls interested in coding and STEM topics. This project, sponsored by NSF, is to create a QCD game that will raise public interest in QCD, especially among K-12 girls, and increase interest in coding among girls. 
Through the immersive framework of interactive gameplay, this QCD phone game will allow the public to peek into the QCD research world. The game design will fall into the ``Match 3'' genre, which typically attracts a higher ratio of female players. The game will be implemented initially as a phone app, and the gameplay would require learning simple QCD rules to progress. By leveraging the willingness of players to engage with the rules of an entertaining game, they are able to easily learn a few principles of physics. The game is now available to download from
the Google Play store (\url{https://play.google.com/store/apps/details?id=com.gellab.quantum3})
and the Apple Appstore (\url{https://itunes.apple.com/gb/app/quantum-3/id1406630529})!

We formed a development team of MSU undergraduate students to make the game and provided them with a QCD curriculum.
The game will be tested at MSU outreach activities, as well as among local K-12 girls through school activities, and feedback will be used to improve the design. The final game can be easily distributed through various app stores and impact will be measured through a follow-up survey. If such a new direction works to attract more girls to coding and physics, one should develop more games to engage more girls in STEM.
}

\FullConference{The 36th Annual International Symposium on Lattice Field Theory - LATTICE2018\\
    22-28 July, 2018\\
    Michigan State University, East Lansing, Michigan, USA.}

\begin{document}

\vspace{-0.5cm}
\section{Introduction}
\vspace{-0.5cm}

\textbf{Motivation: } 
There is a critical need to improve participation of women not only in physics but specifically in computational science. 
As of 2015, women make up just 16 percent of all computer-science graduates~\cite{crn-28-5}, down from 36 percent in 1984~\cite{Gries:1986:TS:6617.6631}.
Nationwide efforts have been launched to get more girls into coding, such as 
``Girls Who Code''~\cite{girls-who-code} and 
CODE~\cite{code}. Many universities are holding physics camps to attract girls into STEM. These are all good efforts to help girls overcome the disadvantages they face
when they are often ``the only woman in the room''. 

Here, we present a complementary approach from the traditional one to improving the gender diversity of the field. According to a recent review article in the New York Times, 
``Some educators now believe gaming could be a way to get girls interested in coding, and even to increase the numbers of girl in STEM classes and schools''~\cite{nyt-girls-coding}. 
Inspired by these recent findings, we are developing a QCD phone game that will appeal to the female demographic.
 
Gaming is a fun and effective gateway for girls to enter the world of physics. The laws of physics are often 
simple and elegant. 
Indeed, they may not be more complicated than some of the very intricate board, card or video games that young people routinely commit to memory. We hope to show that they should not be afraid to pursue another set or rules simply because it is ``physics''. 
QCD is a difficult subject, generally considered abstract and frightening. 
Through the medium of games and play, we hope to open some young minds to the idea that it can be fun. 
By developing a game on the subject of QCD, we can help players avoid shying away from the subject and encourage them to pursue it further instead. 

Putting the game on a portable device makes distribution significantly easier, since most young people carry a phone even if they don't have their own laptop or desktop. For those who do not have a phone, they might be more able to access their parents' phones than a computer.  There are many chances throughout the day while waiting for buses or standing in lines to use an app. The game should contain many levels with simple goals such that one can make progress during short 5 to 10-minute intervals of waiting during the day. Parents would not mind giving their kids ``screen time'' during these periods, since children would be learning while playing.
Many of us have witnessed the power of learning through game-playing on portable devices for spelling, addition and other simple skills. We believe the underlying rules of QCD will be learned just as easily. 

\vspace{-0.5cm}
\section{Design and Implementation}
\vspace{-0.5cm}
\textbf{QCD Phone Game: }
The idea is to design a fun game with simple rules that can reach a younger audience than existing physics apps. 
There are a few nicely done physics apps currently available. 
Take ``Particle Adventure''~\cite{particle-adventure} for example. It is an app that is convenient for people who are interested in physics to look up certain facts and learn more. 
Another game, ``Particle Clicker''~\cite{particle-clicker}, is a much simpler game to play where the role of the player is that of project manager. 
Our design of the QCD game will impose the simple laws of physics as rules and open the door to a younger audience.

Another special focus in the app is to design a game that will be attractive to girls. 
Although much of the gaming industry is dominated by a male demographic, games like ``Candy Crush''~\cite{candy-crush} and ``Bejeweled''~\cite{bejeweled} have much higher female-player participation. 
For this reason we have designed a game of the same genre, ``Match 3''. 
In such games, the player selects tiles on a grid to swap elements. If a row or column of 3 matching tiles is formed, those tiles are removed from the grid and are refilled with new tiles. Additional bonuses are given for forming matches of more than 3, and various challenges are added as the player moves to higher levels. This genre has been extremely popular for over a decade and continues to be one of the most played, with classic games remaining popular and new titles constantly appearing. Players are drawn to the easy-to-grasp rules, simple goals, and scalable challenge level.

\begin{figure}[htbp]
\centering
\includegraphics[width=0.25\textwidth]{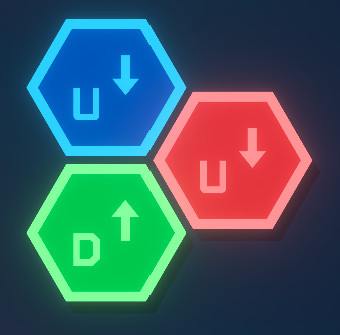}
\includegraphics[width=0.3\textwidth]{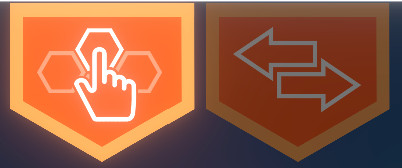}
\includegraphics[width=0.25\textwidth]{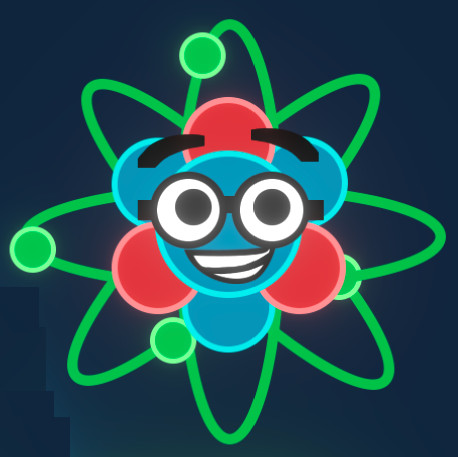}
\vspace{-0.2cm}
\caption{(Left) The quark representation used in the game. (Middle) The icons to activate select and swap modes in game. (Right) Atom, the friendly mascot, gives instructions and advice.}
\label{fig:dof-swap-atom}
\end{figure}

Our physics game adopts similar gameplay on a hexagonal grid, which more naturally represents triangular baryons. Each tile of the grid represents a quark, see the left-hand side of Fig.~\ref{fig:dof-swap-atom}, and the player matches these quarks to form various hadrons. When a hadron is formed, the quarks that make it up disappear and are replaced. The game starts by introducing the central quantum number of the game: color. More concepts are presented to the player, and they learn new rules at each new level. The game will quickly introduce flavor, and as levels advance, antiparticles and spin appear as well. At the higher levels of the game, exotic mesons and baryons such as pentaquarks may be added to Quantum~3 in the future.

\textbf{Game Implementation: }
The goal of this game is to get average K-12 students, especially young girls, playing and learning. This is done through fun gameplay that is intuitive and engaging. That is why we made the core of the game very simple. This game is all about making trios of 1 red, 1 green, and 1 blue tile. It is very easy for the average person to wrap their head around that idea and that core will never change on them, even when things get more complicated.

The mechanics of the game are simple and intuitive. All the player does is press and drag their finger to select tiles. Depending on which of the two modes they are in, they can perform one of two simple actions; see the middle of Fig.~\ref{fig:dof-swap-atom}. If they are in ``Select Mode'', they can select three tiles to form a red-green-blue trio. If they are in ``Swap Mode'', they can select two adjacent tiles to exchange their positions. 

To introduce concepts and teach the player how to play the game, we have Quantum~3's mascot Atom, depicted on the right side of Fig.~\ref{fig:dof-swap-atom}. Atom adds personality to the content and provides a whimsical way to present instructions and guidance throughout the game.

\begin{figure}[htbp]
\centering
\includegraphics[width=0.35\textwidth]{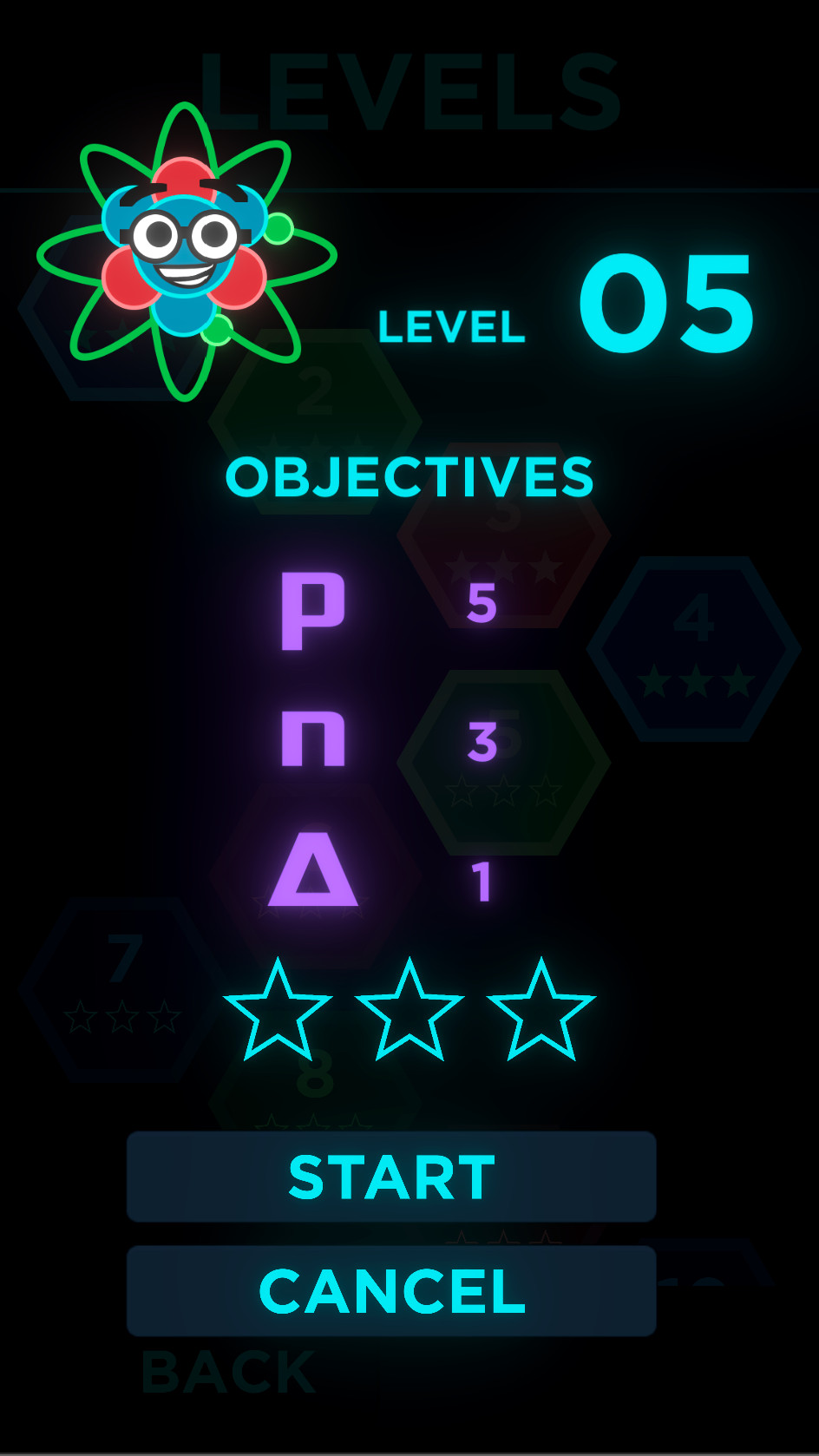}
\includegraphics[width=0.35\textwidth]{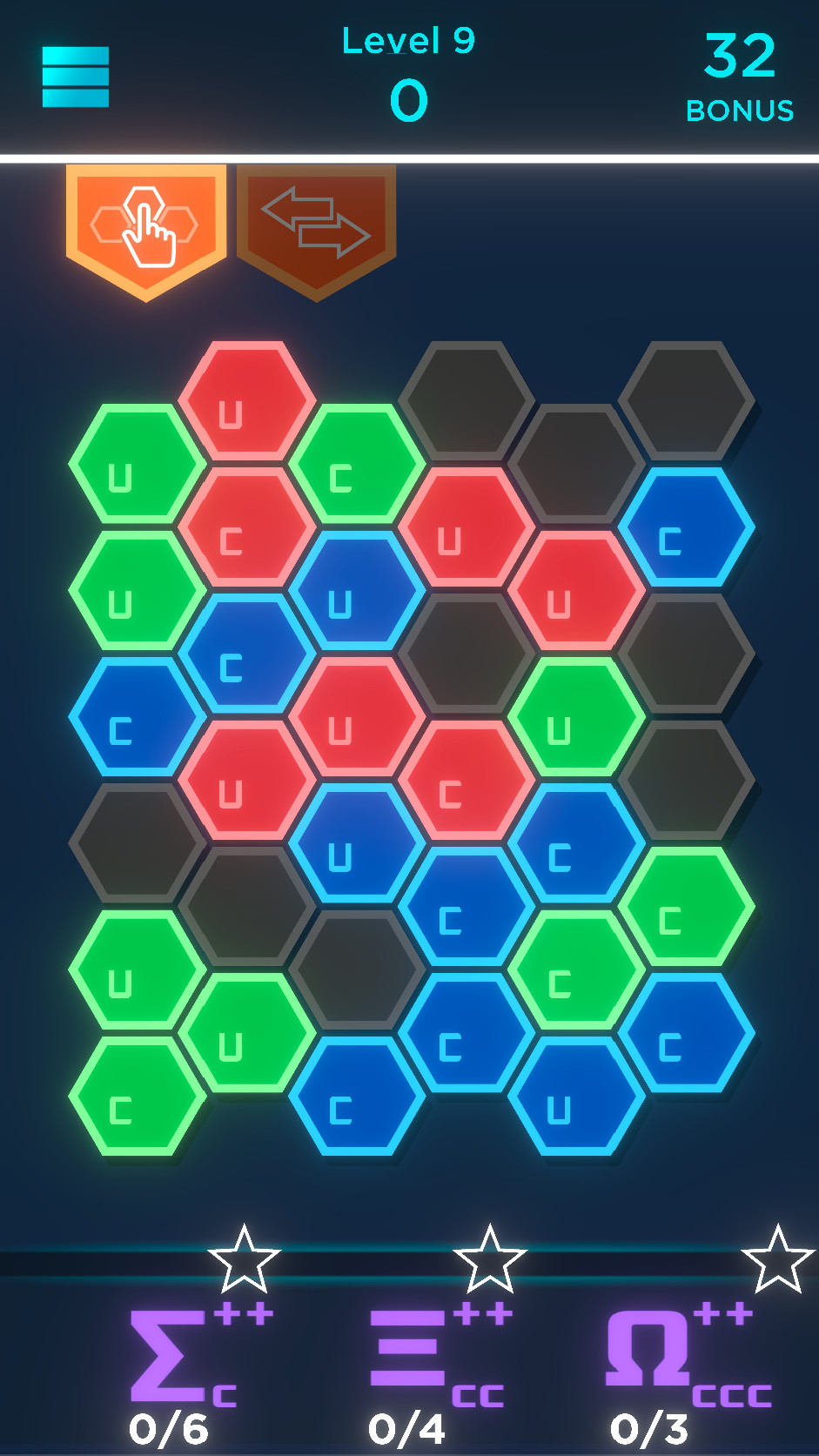}
\vspace{-0.4cm}
\caption{(Left) The level-selection screen shows the objectives and how many stars have been earned on previous attempts. (Right) The playfield is filled with quarks that must be matched to complete the objectives given at the bottom.}
\label{fig:level-field}
\end{figure}

\textbf{Introducing Baryons and Flavor: }
Through the core gameplay of selecting trios, the player makes baryons. We introduce the concept of flavor and that different combinations of three flavors can form different baryons. The first baryons the player will make are the commonly known protons and neutrons so when asked to make baryons with more complicated names, the player will have reference to particles that they already know. To encourage thoughtful action and meaningful play, each level will have the player creating certain baryons as objectives, see the left side of Fig.~\ref{fig:level-field}, to advance in the game. To further encourage the player to think about which baryons are being created, we use a three-star system to rate player efficiency. The player cannot waste moves making the baryons that don't meet objectives or using many swaps if they want to achieve three stars. In playtesting we have seen that kids are highly motivated to replay levels, and thus learn the material better, in order to get all the stars. By providing a system that rewards efficient play but does not block progressing to the next level if a player is doing poorly, we avoid discouraging players that simply aren't as skilled. The three-star system rewards players for doing well, while making it impossible to lose a level.

\begin{figure}[htbp]
\centering
\includegraphics[width=0.6\textwidth]{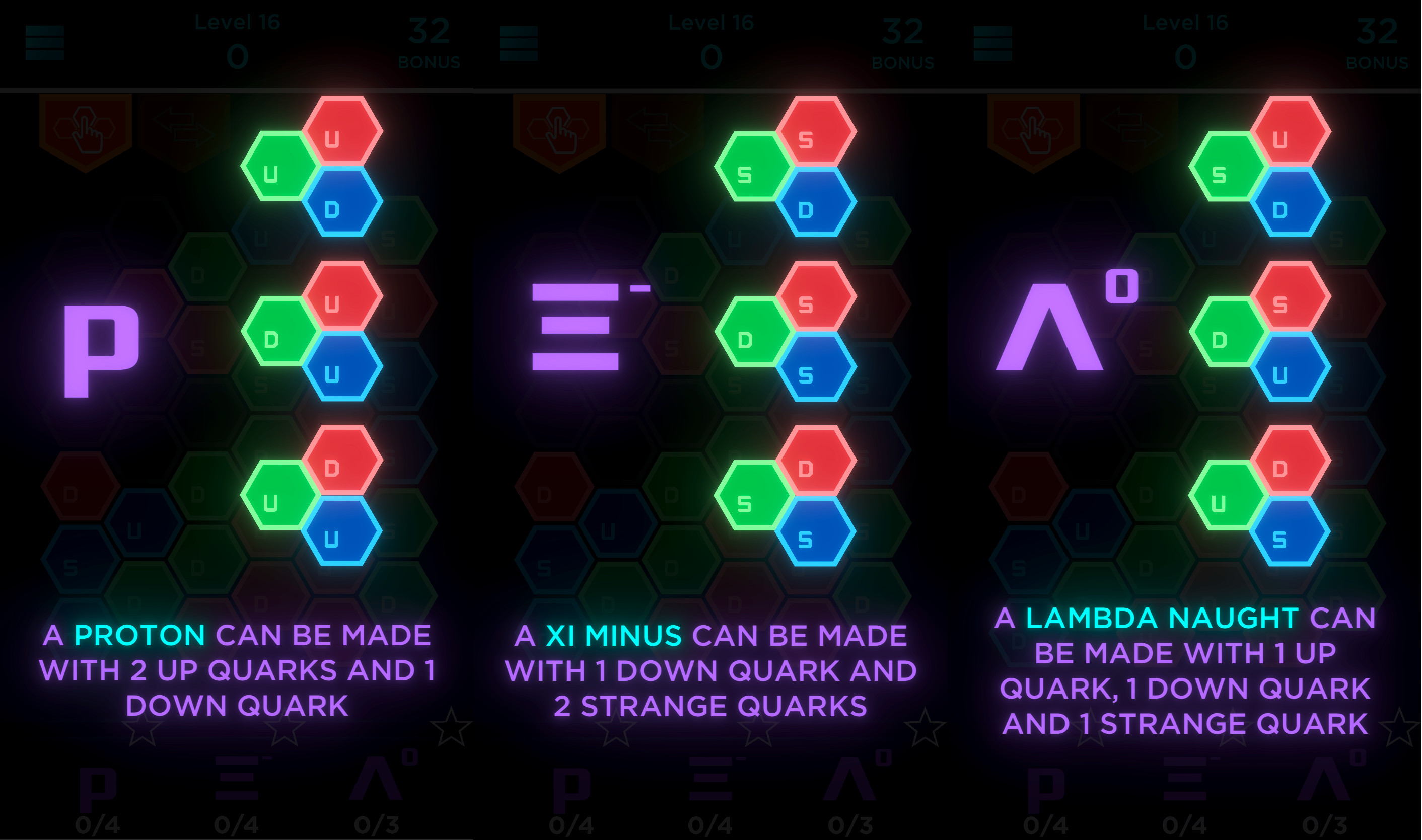}
\vspace{-0.4cm}
\caption{Information panels remind the player how to form the baryons needed for the current level's objectives. Information for all particles could also be made available in a particle index.}
\label{fig:info}
\end{figure}

Since we intend for the players to learn passively by playing the game, we want to avoid pressuring the player to memorize which combinations of flavors make which baryons. Rather, we provide info-panels to the player whenever they want. All they have to do is press an objective, and an info-panel like Fig~\ref{fig:info} will slide into view. Although the player will never need to memorize, playtesting at the Traverse City Film Festival showed that players are not only memorizing the formulae for baryons, but are able to predict how to form baryons they have not previously seen. How? Consistency.

\begin{figure}[htbp]
\centering
\includegraphics[width=0.5\textwidth]{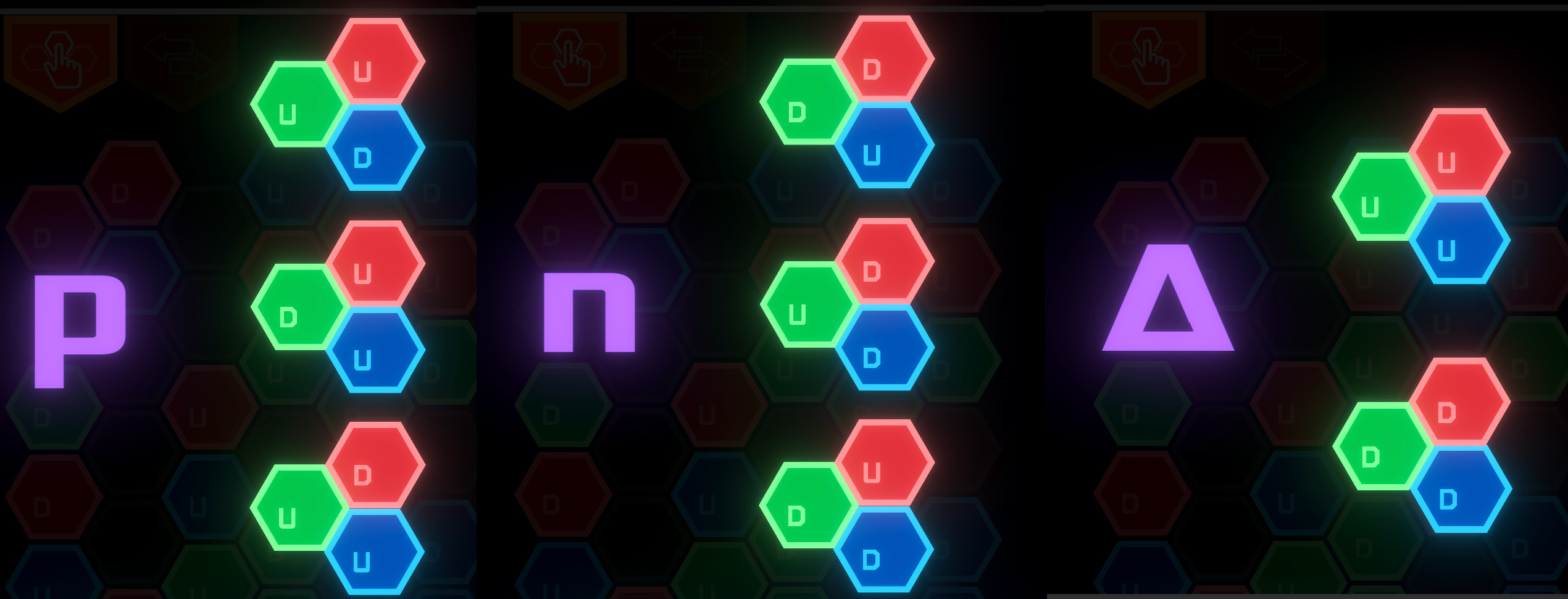}
\includegraphics[width=0.5\textwidth]{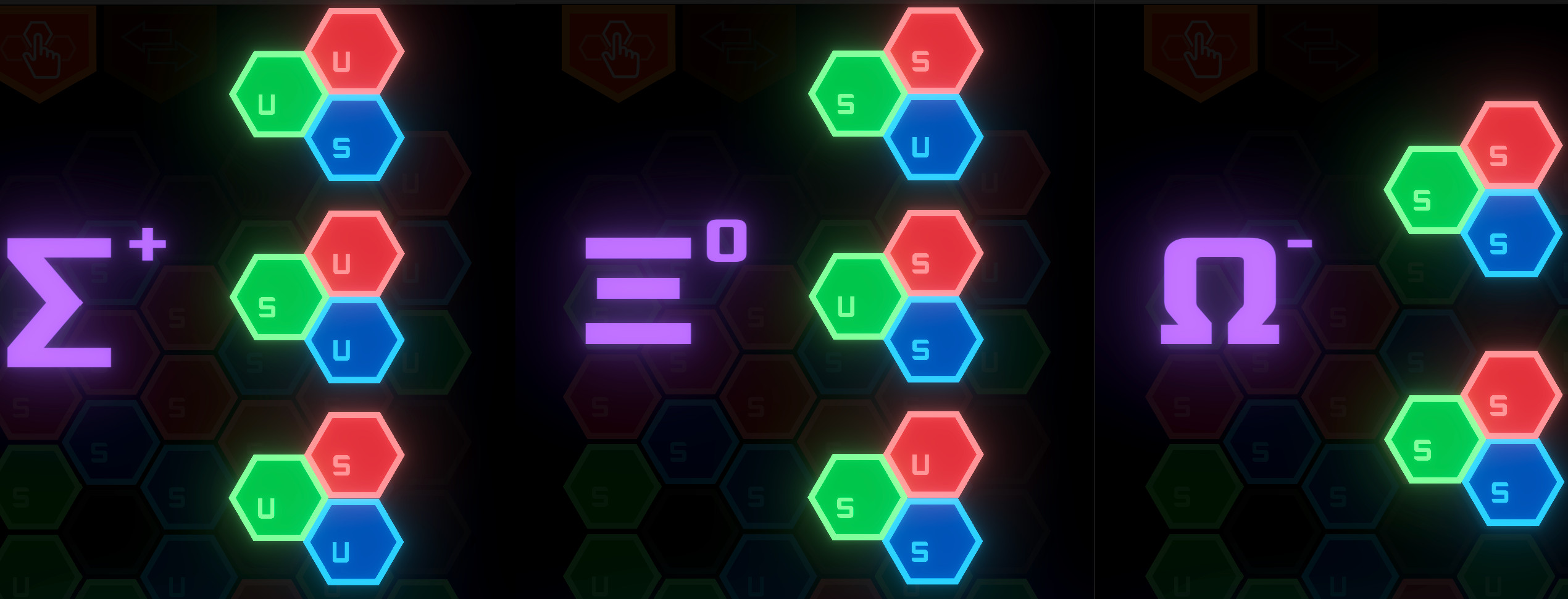}
\vspace{-0.4cm}
\caption{(Top) Objectives for level 6. Note the $\Delta$ is to the right.
(Bottom) Objectives for level 7, which introduces the strange quark. The objectives use one, two and three strange from left to right.}
\label{fig:objectives}
\end{figure}

In game design, consistency is key. It allows the player to get familiar with a system and trust their intuition. Thus, we made sure all levels were set up in a consistent manner. The first two objectives at the bottom of the level are always ordered in a way so that the flavor you were introduced to first takes priority. The order we introduce the flavors in is up, down, strange, charm, and finally bottom. The last objective is usually three of a kind or three different flavors. Because of this setup, we have players who do not even look at the info-panels before making the baryons they need. See, for example, the objectives shown in Fig.~\ref{fig:objectives} where we introduce a new flavor, strange, along with its new baryons. Playtesting shows it is already very intuitive to the player, because the pattern of the objectives is familiar from the previous level.

Through intuitive and consistent design, there will be subtle learning and we avoid heavy-handed teaching. The player will learn the systems of QCD the same way they learn the systems of the game.

\begin{figure}[htbp]
\centering
\begin{tabular}{cc}
\begin{tabular}{c}
\includegraphics[width=0.2\textwidth]{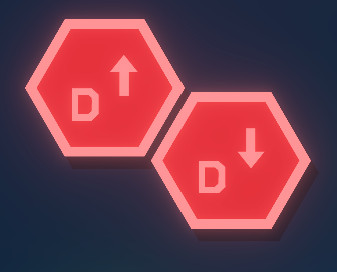}
\includegraphics[width=0.3\textwidth]{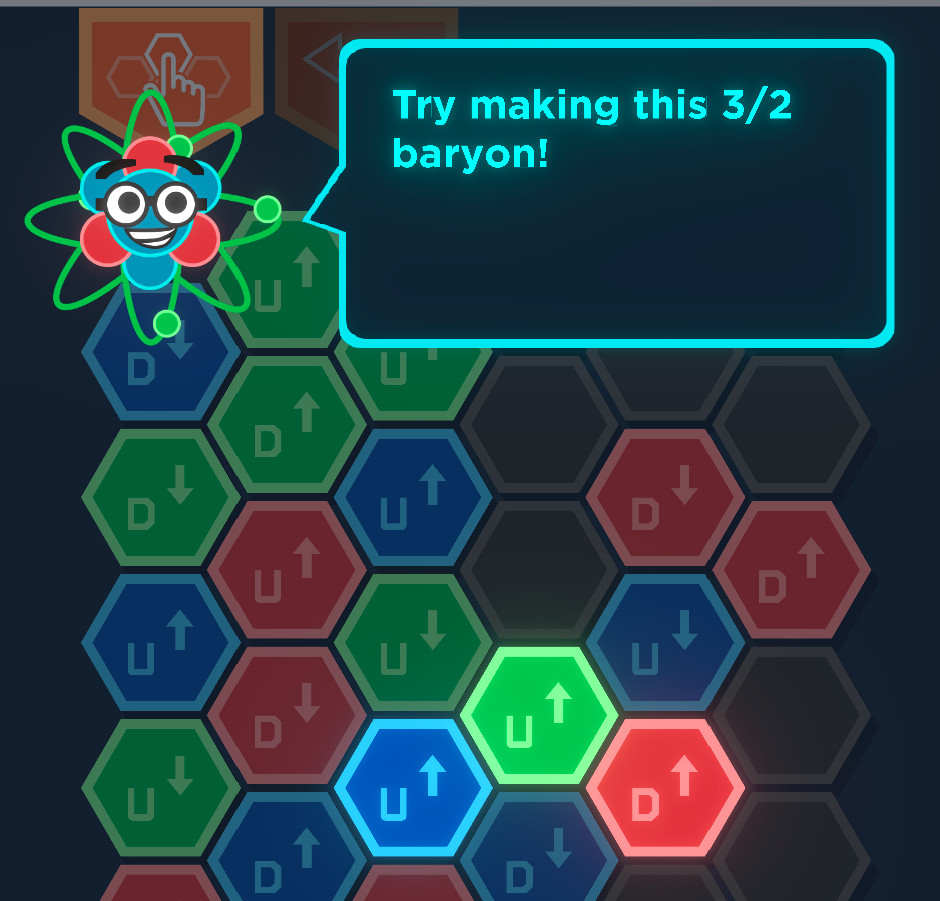}
\end{tabular} &
\begin{tabular}{c}
\includegraphics[width=0.4\textwidth]{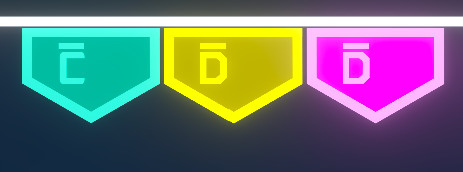} \\ 
\includegraphics[width=0.4\textwidth]{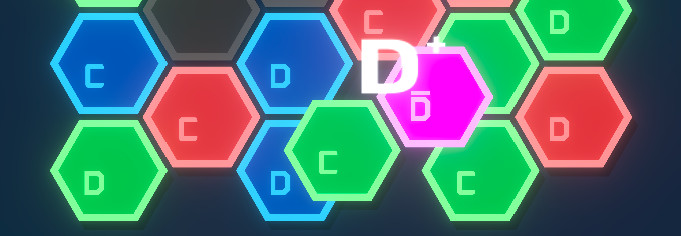}
\end{tabular}
\end{tabular}
\vspace{-0.4cm}
\caption{(Left) The in-game representation of quark spin.
(Middle) Atom shows the player how to make spin-3/2 baryons.
(Upper right) Antiquark power-ups accumulate at the top of the board as the player forms baryons. (Lower right) Forming a meson clears a row from the field, which helps to refresh the field during difficult later levels.
}
\label{fig:spin-antiquark}
\end{figure}

\textbf{Spin: }
Spin complicates things, and it was challenging to implement it without breaking the rules and intuition the player has built through the game up to the point where it is introduced. For the purposes of Quantum~3, we have simplified spin to being represented by either an up or down arrow; see the left-hand side of Fig.~\ref{fig:spin-antiquark}. We then introduce the player to the idea that a baryon with 3 arrows all pointing in the same direction is called a spin-3/2 baryon, otherwise it's a spin-1/2 baryon. While this complicates the objectives, spin is introduced later in the game, where the player should be looking for more of a challenge. It does not inherently break what the player has learned, since they are still making trios of red, green, and blue. See the middle of Fig.~\ref{fig:spin-antiquark}, where Atom explains that 2 up quarks and 1 down quark do not always make a proton, depending on its spin. The levels that incorporate spin also follow consistent level objectives, where the player is given the chance to make a baryon they've already made and also its spin-3/2 version.

\textbf{Mesons, Dead Cells, and Art: }
We always wanted antiquarks and mesons to be represented in the game, but we didn't want to stray from the core gameplay concept of making trios of red, green and blue. The fact that antimatter comes in anticolors and mesons are matches of two instead of three makes the inclusion of antiquarks on the playfield simply too detrimental to the core gameplay. Thankfully, we stumbled upon a beautiful power-up system that lets us include antimatter, see Fig.~\ref{fig:spin-antiquark}, and also makes the game a lot more fun. Every time the player forms a baryon, they have a chance to spawn an antiquark in the upper-right corner of the screen. The player can then choose to drag that antiquark onto a quark on the playfield to form a meson. When a meson is formed, the game, very satisfyingly, clears a row of tiles from the field. Quarks can only combine with antiquarks of the appropriate anticolor, but we do not want the player have to memorize that cyan is antired and so on. Rather, the moment you pick up an antiquark, the quarks available to combine with it are automatically highlighted.

\begin{figure}[htbp]
\centering
\includegraphics[width=0.3\textwidth]{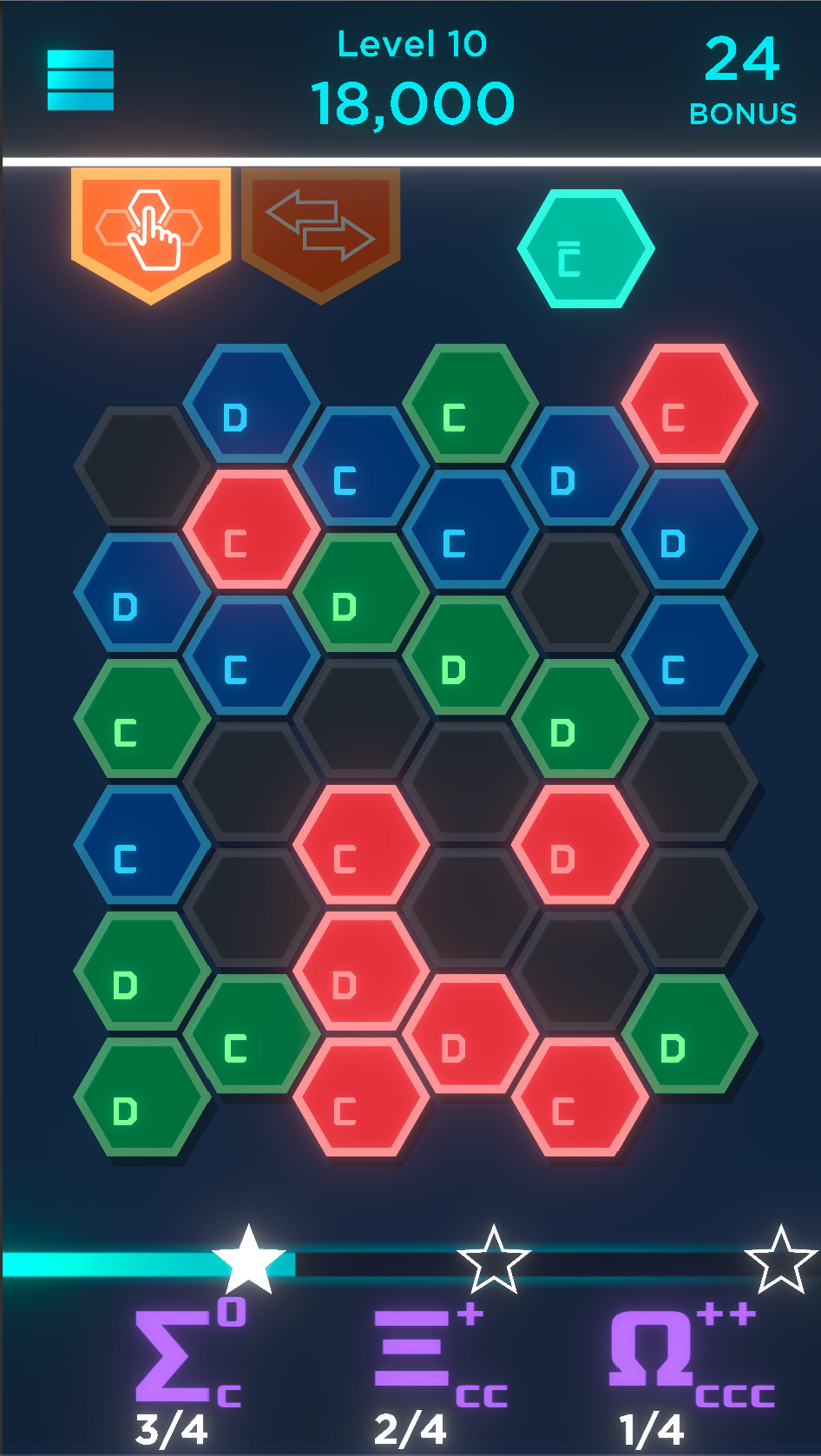}
\includegraphics[width=0.3\textwidth]{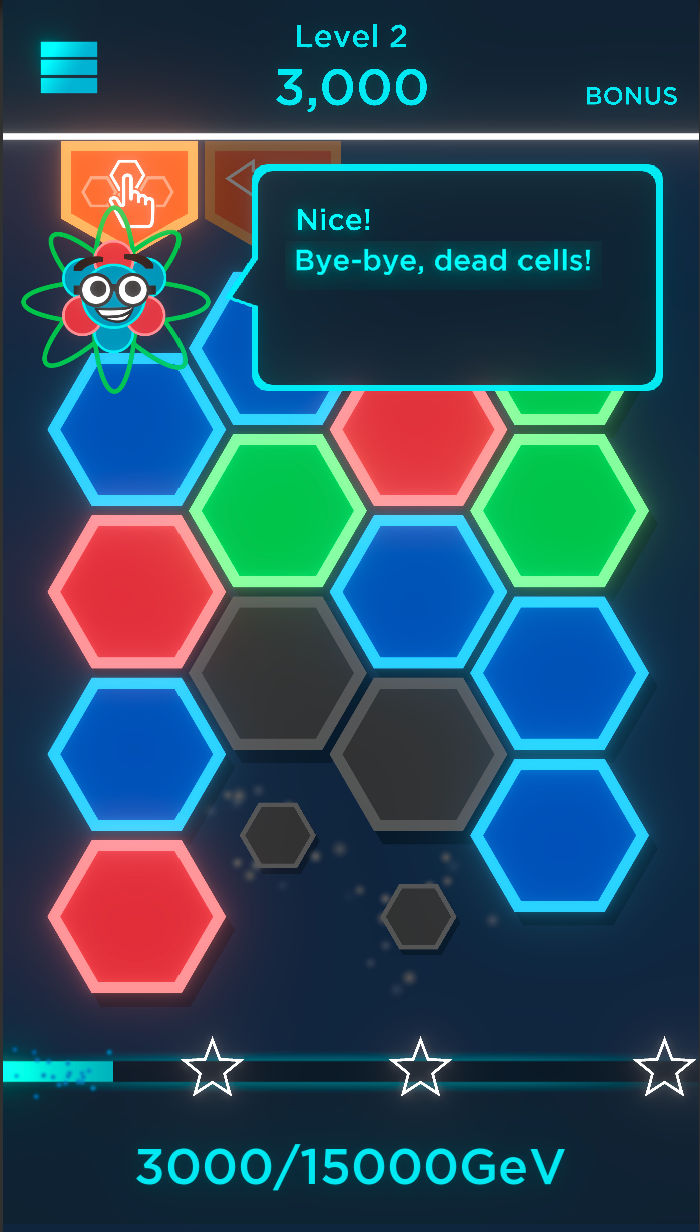}
\vspace{-0.4cm}
\caption{
(Left) Dragging an antiquark onto the board highlights matching-color quarks. 
(Right) Atom introducing dead cells.}
\label{fig:antiquarks-deadcell}
\end{figure}

Dead cells, shown on the right-hand side of Fig.~\ref{fig:antiquarks-deadcell}, were introduced to the game as a way to limit the number of available matches. Unlike other match-3 games that usually have around 6--8 colors, Quantum~3 can only have 3 colors, which makes matches in the initial levels way too plentiful. Dead cells have a chance to fill in the grid just like quarks, but when they get to the bottom of the field, they fall off. In addition to limiting the number of available matches, removing dead cells turns out to be one of the most enjoyable parts of the game for our playtesters. It is immensely satisfying to get rid of a large group of dead cells with a smart move, especially when you combine it with the meson mechanic.


Many things went into making this game fun and well designed, while keeping intact the integrity and accuracy of the physics. One of the last things that should be mentioned is the art. The quark art was designed to hold all the necessary information while being as uncrowded as possible. We wanted things to be clean and unbusy, with mostly flat, succinct colors. The level layout took many iterations to improve usability. We needed the board to be as large as possible, but also have the objectives noticeable and the modes within reach. All this helps a great deal in keeping players from getting frustrated with the game.

\vspace{-0.5cm}
\section{Release and Outlook}
\vspace{-0.5cm}
Quantum 3 will be released to the App Store in the near future and will hopefully become available on Android and Steam as well. Through analytics, we will gather data from players who download the game and continue development.

\vspace{-0.5cm}
\section*{Acknowledgments}
\vspace{-0.5cm}
This project is supported by the US National Science Foundation under grant PHY 1653405 ``CAREER: Constraining Parton Distribution Functions for New-Physics Searches''.
We thank the Games for Entertainment and Learning Laboratory at Michigan State University for providing the infrastructure to develop this game. 
We thank the rest of the GEL team, Harrison Sanders, Rebecca Roman, Roman Firestone, and Colleen Little, who worked hard to develop this game.
Special thanks to Saul Cohen and Esther Cohen-Lin for their suggestions and playtesting in the early development stages. 
Thanks to everyone who participated in our survey during the Lattice 2018 conference and Traverse City Film Festival. 

\vspace{-0.5cm}
\bibliographystyle{apsrev}

\end{document}